\documentclass[twocolumn]{jpsj3}
\usepackage{graphicx}
\usepackage{color}
\usepackage{bm}
\newcommand{\nn}{\nonumber}

\def\sgn{\mathrm {sgn}}
\def\diag{\mathrm {diag}}

\title{Zero Energy Modes and Statistics of Vortices in Spinful Chiral $p$-Wave Superfluids}

\author{Takuto \textsc{Kawakami}\thanks{E-mail address: kawakami@mp.okayama-u.ac.jp}, Takeshi \textsc{Mizushima}, and Kazushige \textsc{Machida}
}
\inst{Department of Physics, Okayama University, Okayama 700-8530 
}
\abst{
	  The possible stable singular vortex (SV) and half-quantum vortex (HQV) 
	  of the superfluid $^3$He-A phase confined in restricted geometries are investigated.
	  The associated low-energy excitations are calculated in connection with the possible existence of Majorana zero modes
	  obeying non-Abelian statistics.
	  The energetics between those vortices is carefully examined using the standard Ginzburg-Landau (GL) functional 
	  with a strong-coupling correction.
	  The Fermi liquid effect, which is not included in the GL functional, is considered approximately within the London approach.
	  This allows us to determine the stability regions in pressure, temperature, and applied field for SV and HQV.
	  The existence of the Majorana zero mode and its statistics, either Abelian or non-Abelian under braiding of SVs,
	  is studied by solving the Bogoliubov-de Gennes equation for spinful chiral $p$-wave superfluids 
	  at sufficiently low temperatures.
	  We determined several conditions controllable external parameters 
	  for realizing the non-Abelian statistics of Majorana zero modes e.g., pressure, field direction, and strength.
	  }
\kword{chiral $p$-wave superfluids, vortex, zero energy state, non-Abelian statistics, Ginzburg-Landau theory, Bogoliubov-de Gennes equation}
\begin{document}
\maketitle
\section{Introduction}
Much attention has been paid to various exotic vortices and the associated low-energy excitations 
in spin-triplet $p$-wave superfluids and superconductors.~\cite{leggett, salomaaRMP, maenoRMP}
In particular, there ha{s} been a considerable number of investigations for the quasiparticle 
whose energy eigenvalue is exactly zero and whose creation and annihilation operators are self-Hermitian: $\gamma_{E=0}=\gamma_{E=0}^\dagger$.~\cite{moore,read}
Fermionic field operators that have self-Hermitian creation and annihilation operators are called the Majorana fermions~\cite{majorana}, 
and their quasiparticle energy modes with $\gamma_{E=0}=\gamma_{E=0}^\dagger$ are called the Majorana zero-energy states (ZESs).
The Majorana ZESs localized at vortices are considered useful for fault-tolerant quantum computations because they can obey the non-Abelian statistics.\cite{read, ivanov, stern, stone, nayak}

The localized Majorana ZESs have been pointed out in the quasi-hole of the Pfaffian state of the quantum Hall state with the 5/2 filling.~\cite{moore,read}
The other candidate systems that support Majorana ZESs are 
quantized vortices in chiral $p$-wave superfluids or superconductors~\cite{ivanov, gurariePRB, gurarieAnn, mizushimaPRL, tsutsumiPRL, mizushimaPRA, tsutsumiJPSJ, massignan}, 
the surface Andreev bound state of chiral or time reversal symmetric $p$-wave superfluids~\cite{stone2, qi, zhang, volovikB, nagato, tsutsumiPre},
the junction between an $s$-wave superconductor and a topological insulator~\cite{fu,tanaka, linder, chamon}, 
and the $s$-wave superfluid with particular spin-orbit interactions~\cite{sato}.

First, we consider Majorana ZESs bound at a singular vortex in spin-polarized $p$-wave superfluids, which involves the particle-hole symmetry $\gamma_{E}=\gamma_{-E}^\dagger$.
Then the existence of the exact ZES is guaranteed by the index theorem~\cite{tewari} and the analytic solution of the Caroli-de Gennes-Matricon state
when the vorticity is odd.~\cite{gurariePRB, mizushimaPRA, kopnin, volovikbook1}
Therefore, the existence of the Majorana ZESs is topologically protected.
The eigenstates of the system constructed by Majorana quasiparticles are described by the occupation of the complex fermion state, 
that is, a linear combination of two Majorana quasiparticles, $c_{2j} = (\gamma_{2j-1} + i\gamma_{2j})/\sqrt{2}$.
For spin-polarized systems, a complex fermion is formed by a pair of the Majorana ZESs $\gamma_j$ and $\gamma_{j+1}$ in between spatially separated vortices,
giving rise to the nonlocality of the complex fermion.
If a well-isolated vortex has an odd number of Majorana ZESs, the complex fermion state is necessarily constructed by the spatially separated Majorana ZES and the vortices obey the non-Abelian statistics because of the nonlocality of their excitations.~\cite{ivanov}
For the system where the spin degrees of freedom, for example, the spin $\left|\uparrow \uparrow \right>$ and $\left|\downarrow \downarrow \right>$ pairs, survive,
an isolated singular vortex (SV) has a degeneracy of the ZESs in two spin sectors.~\cite{kawakami1}
The statistics of the SV in spinful superfluids depends on the ways of forming a complex fermion as follows.
If the two Majorana quasiparticles constructing a complex fermion eigenstate are in a single vortex core, the statistics is Abelian. 
If they are in spatially separated vortices, however, the statistics is non-Abelian.
Thus, a half-quantum vortex (HQV), that has a singularity in either 
the spin $\left|\uparrow \uparrow \right>$ or $\left|\downarrow \downarrow \right>$ component of the order parameter,
is more feasible for the non-Abelian statistics.~\cite{ivanov}

The superfluid $^3$He-A phase is one of the most effective candidates of the $p$-wave superfluid associated with the Majorana ZES
in the sense that the ground states are most established~\cite{vollhardt, leggett} and there are many experiments and samples with controllable parameters.~\cite{wheatley, greywall, yamashita, saunders, kono}
Since the superfluid $^3$He-A phase is a spinful system,
the statistics of the vortex in spinful $p$-wave superfluids should be clarified.

In order to observe the Majorana character, it is necessary to set up a two-dimensional system
where the thickness of the sample is much less than the dipole coherence length.
A possible experiment has been set up by Saunders group,
where they confine superfluid $^3$He in a slab geometry whose thickness is 0.6 $\mu$m under the following conditions:
The pressures are between 0 and 5.5 bar, the temperature is cooled to 350 $\mu$K, 
and the external field is applied perpendicular to the slab to carry out the NMR observation.
Then the group observes the phase transition from the A phase to an unknown phase
in a low-pressure and low-temperature region.
Investigation of vortices in spinful chiral $p$-wave superfluids and their features will also lead 
us to understand such a characteristic of $^3$He-A in a slab in future. 

For rotating experiments, Yamashita $et$ $al$.~\cite{yamashita} have recently performed an experiment on parallel-plate geometry 
intended to observe the HQV in superfluid $^3$He-A.
The superfluid is confined in a cylindrical region with a radius $R=1.5$ mm
and a height of 12.5 $\mu$m, sandwiched between parallel plates.
A magnetic field $H=26.7$ mT ($\parallel$$\bm{z}$) is applied
perpendicular to the parallel plates, and the pressure $P$=3.05 MPa. 
At this pressure, the strong-coupling effect due to spin fluctuations becomes important.~\cite{anderson}
Since the gap of 12.5 $\mu$m
between plates is small compared with the dipole coherence length
$\xi_d\sim 10 \mu$m, the $l$-vectors, which indicate the direction of the orbital 
angular momentum of Cooper pairs, are always perpendicular to the plates.
Also the $d$-vectors are confined within the plane by the applied field $\bm{H} \parallel \bm{z} $
because the dipole magnetic field is $H_d\sim 2.0 $ mT~\cite{leggett}, where $H$ tends to align the
$d$-vectors perpendicular to the field direction.
Yamashita $et$ $al$. conducted an investigation in various parameter spaces, such as
the temperature $T$, or the rotation speed $\Omega$ up to $\Omega=6.28$ rad/s
using the rotating cryostat in ISSP, University of Tokyo, capable of achieving a maximum rotation speed of $\sim$12 rad/s;
however, there is as yet no evidence of the HQV~\cite{yamashita}.

The stability of HQVs has been argued since HQVs were pointed out by Volovik and Mineev~\cite{mineev}.
The hydrodynamical calculation taking account of the Fermi liquid (FL) correction shows that the HQV is energetically stable against the SV.~\cite{cross, salomaahqv, salomaaRMP, chung, vakaryuk}
However, in our previous study~\cite{kawakami2}, we have carried out the calculation on the basis of the full Ginzburg-Landau (GL) theory 
taking account of the strong-coupling correction due to spin fluctuations, 
which play a crucial role in the stability of the A phase at high pressures, 
without the FL correction.
This implies that the HQV is unstable unless the pairing phase becomes the so-called A$_2$ phase under a strong external field.~\cite{kawakami2}
In this study, we find that the contributions of the strong-coupling and FL effect to the stability of the HQV are competitive 
at realistic rotation speed and high pressure in the experiment using the rotating cryostat in ISSP.

Therefore, in order to clarify the statistics of vortices in the spinful $p$-wave superfluid, 
we have to examine the statistics of the SV, in addition to that of the HQV on an equal footing.
The argument on the statistics of vortices is valid in adiabatic and quantum limits, where the ZESs are energetically distinct from the other core bound states with finite energies. This implies the absence of the decoherence of ZESs and requires the temperature to be sufficiently lower than the energy difference between core bound states $T^2_c/T_F $ where $T_c$ is the transition temperature and $T_F$ is the Fermi temperature.

This paper is arranged as follows:
In \textsection\ref{formulation}, we derive the spinful Bogoliubov-de Gennes (BdG) equation and introduce the GL free energy.
In \textsection\ref{energetics}, we discuss the energetics of the vortex textures and the stability of the HQV.
We calculate the energetic advantage of the HQV originating from the FL correction
and the disadvantage originating from the spin-fluctuation strong-coupling correction
using both the London limit calculation and the full GL calculation.
In \textsection\ref{SV}, we examine the statistics of the SV using numerical calculation with the spinful BdG equation.
In \textsection\ref{Hperp}, we discuss the case in which the direction of $d$-vectors is perpendicular to the external field.
In this section, we clarify that when the vortex distance is finite, two eigenstates originating from ZESs in different spin sectors
do not hybridize with each other so that the braiding of the SV does not commute necessarily.
In \textsection\ref{Htilt}, we consider the case where the direction of $d$-vectors is tilted from the direction perpendicular to the external field.
In this situation, we demonstrate that the Zeeman effect due to an external field parallel to the $d$-vectors causes the hybridization of the two Majorana ZESs.
In the final section, we present our summary and conclusions.

\section{Formulation}\label{formulation}
\subsection{Spinful Bogoliubov-de Gennes equation}
In general the OPs for spin triplet superfluids are described as 
\begin{eqnarray}\label{op}
	\hat{\Delta}(\bm{r}_1, \bm{r}_2) \equiv
	\left[\begin{array}{cc}
	\Delta_{\uparrow \uparrow } (\bm{r}_1, \bm{r}_2)   & \Delta_{\uparrow   \downarrow } (\bm{r}_1, \bm{r}_2) \\
	\Delta_{\downarrow \uparrow } (\bm{r}_1, \bm{r}_2) & \Delta_{\downarrow \downarrow } (\bm{r}_1, \bm{r}_2)
	\end{array}\right],
\end{eqnarray}
where $\Delta_{\sigma \sigma'}=-V(\bm{r}_1, \bm{r}_2)\left< \psi_{\sigma }(\bm{r}_1)\psi_{\sigma'}(\bm{r}_2)\right>$ and $\psi_{\sigma}(\bm{r})$ is the field operator of fermions
with spin $\sigma=\uparrow,\downarrow$.
The spinful mean-field Hamiltonian of the spin-triplet superfluids and superconductors is described using this notation as
\begin{eqnarray}\label{MFH}
	\mathcal{H} = E_0 + \frac{1}{2} \int d \bm{r}_1 \int d \bm{r}_2 \bm{\Psi}^\dagger (\bm{r}_1)
			\underline{\mathcal{K}}(\bm{r}_1, \bm{r}_2)
			\bm{\Psi} (\bm{r}_2), 
\end{eqnarray}
\begin{eqnarray}
	&\bm{\Psi}(\bm{r}) = \Bigl[\psi_{\uparrow }(\bm{r}), 
	\psi_{\downarrow }(\bm{r}), \psi_{\uparrow }^\dagger(\bm{r}), \psi_{\downarrow }^\dagger(\bm{r}) \Bigr] ^T, &\\
	&\underline{\mathcal{K}}(\bm{r}_1, \bm{r}_2)=
	\left[\begin{array}{cc}
		\hat{H}_0(\bm{r}_1,\bm{r}_2)  & \hat{\Delta}(\bm{r}_1, \bm{r}_2) \\
	-\hat{\Delta }^*(\bm{r}_1, \bm{r}_2)          & -\hat{H}_0^*(\bm{r}_1,\bm{r}_2)
	\end{array}\right], & \\
	&\hat{H}_0(\bm{r}_1,\bm{r}_2) =
	\delta(\bm{r}_1-\bm{r}_2)[H_0(\bm{r}_1) \hat{1} + \mu_n \bm{H}\cdot \hat{\bm{\sigma}}], 
\end{eqnarray}
where $\bm{\Psi} (\bm{r})$ is the spinor in the Nambu space, $\hat{1}$ is a $2\times 2$ unit matrix, 
and $\mu_n$, $\bm{H}$, and $\hat{\bm{\sigma }}$ are the magnetic moment of $^3$He atoms, the external field, and the $2\times 2$ Pauli matrices respectively.
Here, we assume that the superfluids and superconductors are confined by the potential with a magnitude $V_0 \gg |\Delta_{\sigma \sigma'}|$.
Then the single particle part $H_0(\bm{r})$ is given as 
\begin{eqnarray}
	H_0(\bm{r})=
		-\frac{\nabla  ^2}{2M}+V_0\theta(r-R)-\mu
		+i\bm{\Omega} \cdot \left(\bm{r}\times\bm{\nabla}\right),
\end{eqnarray}
where $M$, $\mu$, and $\Omega$ are the mass of the particle, the chemical potential, and the external rotation, respectively.
We set $\hbar=1$.

In general, the OP of $p$-wave superfluids is expanded in terms of the eigenstate of the orbital angular momentum of the Cooper pair $l_z = -1,0,1$.
We assume the system to be spinful chiral $p$-wave superfluids, namely, the orbital ferromagnetic state, so that the only $l_z=-1$ component survives.
This can be realized in a parallel-plate geometry and a slab, where the dipole coherence length is much larger than the thickness of the sample.
Then the explicit expression of the OPs is given as
\begin{eqnarray}\label{op-Rk}
\nn	\Delta_{\sigma \sigma'}\left(\frac{\bm{r}_1+\bm{r}_1}{2},\tilde{\bm{k}}\right) = -A_{\sigma \sigma',-1} \left( \frac{\bm{r}_1+\bm{r}_2}{2}\right)\\
																	\times \frac{\tilde{k}_x-i\tilde{k}_y}{k_F}\exp[-(\tilde{k}^2-k_F^2)\xi_p^2],
\end{eqnarray}
where $\tilde{\bm{k}}$ is the relative momentum and $\xi_p$ is the size of the Cooper pair.
The components in eq.~(\ref{op}) are obtained as the Fourier transformation of eq.~(\ref{op-Rk}) with respect to the relative coordinate
\begin{eqnarray}\label{non-local-op}
\nn	\Delta_{\sigma \sigma'}(\bm{r}_1, \bm{r}_2) = -A_{\sigma \sigma',-1} \left( \frac{\bm{r}_1+\bm{r}_2}{2}\right) \\
								\times \frac{ix_{12}+y_{12}}{8\pi \xi _p^4 k_F}\exp\left[ -\frac{r_{12}^2}{4\xi_p^2}+k_F^2\xi_p^2\right],
\end{eqnarray}
where $k_F$ is the fermi wave number.

We carry out the Bogoliubov transformation from the fermionic field operator to the quasiparticle basis with
$\bm{\Gamma }_\nu = [ \Gamma_{\nu \uparrow },\ \Gamma_{\nu \downarrow },\ \Gamma^\dagger_{\nu \uparrow },\ \Gamma_{\nu \downarrow }^\dagger ]^T$
in the Nambu space:
\begin{eqnarray}
\nn	&\bm{\Psi} (\bm{r}) = \sum\limits_\nu \underline{u}_\nu (\bm{r})\bm{\Gamma }_\nu,&\\
\nn	&\underline{u}_\nu (\bm{r}) =
	\left[ 	\bm{u}_{\nu}^{(1)},\ \bm{u}_{\nu}^{(2)},\ 
	\left\{ \underline{\tau}_1 \bm{u}_{\nu}^{(1)} \right\}^*,\ \left\{\underline{\tau}_1 \bm{u}_{\nu}^{(2)}\right\}^*
	\right],&
\end{eqnarray}
where $\underline{\tau}_i$ is the $4\times 4$ matrix defined by a Pauli matrix.
The transformation matrix $\underline{u}_\nu (\bm{r})$ must satisfy the orthonormality 
$\int d\bm{r}_1 \underline{u}_\nu ^\dagger(\bm{r}_1) \underline{u}_\mu (\bm{r}_1) = \delta_{\nu, \mu } \underline{1}$
and the completeness $\sum_\nu d\bm{r}_1 \underline{u}_\nu (\bm{r}_1) \underline{u}_\nu ^\dagger(\bm{r}_2) = \delta(\bm{r}_1-\bm{r}_2 ) \underline{1}$,
where $\underline{1}=\diag(\hat{1},\hat{1})$ is a $4\times 4$ unit matrix.
In order to obtain the operators of the quasiparticles that diagonalize the mean field Hamiltonian eq.~(\ref{MFH}) as
\begin{eqnarray}
			&\mathcal{H} = E_0+ \frac{1}{2}\bm{\Gamma }_\nu ^\dagger \underline{E}_\nu \bm{\Gamma }_\nu , &\\
			&\underline{E}_\nu \equiv \diag \left(E_\nu^{(\uparrow)},E_\nu^{(\downarrow)},-E_\nu^{(\uparrow)},-E_\nu^{(\downarrow)}\right) ,&
\end{eqnarray}
one can find that the wave functions of the quasiparticles $\bm{u}_{\nu}^{(1)}(\bm{r})$ and $\bm{u}_{\nu}^{(2)}(\bm{r})$ should obey the same BdG equation described as
\begin{eqnarray}\label{egn-eq}\label{bdg}
	&&\int d\bm{r}' \underline{\mathcal{K}}(\bm{r}, \bm{r}')
	\bm{u}_{\nu}(\bm{r'})
	= E_\nu^\sigma \bm{u}_{\nu}(\bm{r}),\\
	&&\bm{u}_\nu (\bm{r}) =
	\left[ 
	u_{\nu}^{\uparrow}(\bm{r}),\ u_{\nu}^{\downarrow}(\bm{r}),\ 
	v_{\nu}^{\uparrow}(\bm{r}),\ v_{\nu}^{\downarrow}(\bm{r})
	\right].
\end{eqnarray}
Then the annihilation operator of the quasiparticle is expressed as
\begin{eqnarray}
	\bm{\Gamma}_{\nu \sigma }= \int d \bm{r} \left\{\bm{u}_{\nu}(\bm{r})\right\}^\dagger \bm{\Psi}(\bm{r}).
\end{eqnarray}

We numerically diagonalize the BdG equation~(\ref{bdg}) under the gap potential given in eq.~(\ref{op-Rk}) with 
\begin{eqnarray}\label{gap-potential}
	A_{\sigma \sigma',-1}(\bm{r}) = A_{\sigma \sigma',-1}^0 \prod\limits_{j=1}\limits^{N_v} \exp[i\kappa_j\theta_j]\tanh\left( \frac{r_j}{\xi_{\sigma \sigma'}}\right),
\end{eqnarray}
where $\kappa_j$ is the winding number of the component in the $j$-th vortex, $\theta_j$ and $r_j$ are the azimuthal angle and radius centered by $j$th vortex, respectively, 
and $\xi_{\sigma \sigma'}$ is the coherence length of the OP in the spin sector $\left|\sigma\sigma'\right>$.
We assume the uniformity of the OPs along the $z$-direction.
For the quasiparticle eigenstate in the BdG equation~(\ref{bdg}), 
we impose the periodic boundary condition with the wave number $k_z$,
that is, $\bm{u}_\nu=\bm{u}_{E,k_z}(x,y)\exp[ik_z z]$ and $\Gamma_{\nu \sigma} = \Gamma_{E, k_z, \sigma} $.
Then the BdG equation~(\ref{bdg}) is block-diagonalized in terms of $k_z$.
In the subspace, the particle-hole symmetry $\{\underline{\tau}_x\underline{\mathcal{K}}\ \underline{\tau}_x\}^*=-\underline{\mathcal{K}}$ gives $\underline{\hat{\tau }}_x \{\bm{u}_{E, k_z}\exp[ik_zz]\}^* = \bm{u}_{-E, k_z}\exp[ik_zz]$ 
and the inversion symmetry along the $z$-direction gives $\bm{u}_{E,k_z}= \bm{u}_{E,-k_z}$ and $E(-k_z)=E(k_z)$. 
Thus, one finds $\Gamma_{E,k_z,\sigma }^\dagger = \Gamma_{-E, -k_z, \sigma }$, 
which implies that the quasiparticle arising from $k_z= 0$ can be the Majorana zero mode 
$\Gamma_{0, 0, \sigma}^\dagger=\Gamma_{0, 0, \sigma}=\gamma^\sigma$.
We focus on the eigenstate with $k_z=0$ throughout this work. 
The numerical diagonalization is carried out by the discrete variable representation method.~\cite{light,baye,manolopoulos,mizushimaPre}

\subsection{Ginzburg-Landau framework}
We use the GL framework to discuss the stable textures and the energetics of the vortex,
which is quantitatively reliable for $^3$He.
Then we assume that the OPs are decomposed to the center-of-mass coordinate
and the orbital degrees of freedom with a relative momentum around $|k|\simeq k_F$.
These components are described with the $3\times 3$ matrix $A_{\mu i}=A_{\mu i}(\bm{r})$ as
\begin{eqnarray}
	\hat{\Delta}(\bm{r}, \hat{\bm{k}}) 
	&=&\left[\begin{array}{cc}
	-A_{xi}+iA_{yi}    & \sqrt{2}A_{z i}  \\
	\sqrt{2}A_{zi}  & A_{xi}+iA_{yi}
	\end{array}\right]\hat{k}_i\\
	&=&\left[\begin{array}{cc}
	 A_{\uparrow \uparrow m} &  A_{\uparrow \downarrow m}  \\
	 A_{\uparrow \downarrow m}  & A_{\downarrow \downarrow m}
	\end{array}\right]\hat{k}_m,
\end{eqnarray}
where $i=x,y,z$, $m=-1,0,+1$, $\hat{k}_\pm = (\hat{k}_x\mp i \hat{k}_y)\sqrt{2}$, and $\hat{\bm{k}}$ is the unit vector 
oriented to the direction of the momentum on the Fermi surface. 
The GL free-energy functional, which is invariant under gauge transformation and under spin and orbital space rotation, is 
well-established~\cite{leggett,vollhardt,salomaaRMP,wheatley,sauls,greywall,thuneberg,fetter,kita} and given by the standard form
\begin{eqnarray} \label{glfun}
		f_{\mathrm{total}}=f_{\mathrm{bulk}}^{(2)}+f_{\mathrm{bulk}}^{(4)}+f_{\mathrm{grad}}+f_{\mathrm{dipole}}+f_{\mathrm{field}},
\end{eqnarray}
\begin{eqnarray}\label{bulk-2-e}
	&f_{\mathrm{bulk}}^{(2)}  =  -\alpha   A_{\mu i}^*A_{\mu i}, &\\
&\nn	f_{\mathrm{bulk}}^{(4)}  =  \beta _1 A_{\mu i}^*A_{\mu i}^*A_{\nu j}A_{\nu j} 
					   + \beta _2 A_{\mu i}^*A_{\nu j}^*A_{\mu i}A_{\nu j} &\\
	& \nn               + \beta _3 A_{\mu i}^*A_{\nu i}^*A_{\mu j}A_{\nu j} 
					   + \beta _4 A_{\mu i}^*A_{\nu j}^*A_{\mu j}A_{\nu i} &\\
	&                   + \beta _5 A_{\mu i}^*A_{\mu j}^*A_{\nu i}A_{\nu j},\label{bulk-4-e} &\\ 
	&\nn	f_{\mathrm{grad}} = K_1(\partial _{i}^*A_{\mu j}^*)
		(\partial _{i}A_{\mu j}) + K_2(\partial _{i}^*A_{\mu j}^*)(\partial _{j}A_{\mu i}) &\\
	& 		 + K_3(\partial _{i}^*A_{\mu i}^*)(\partial _{j}A_{\mu j}), \label{grad-e}&\\
	& f_{\mathrm{dipole}} = g_\mathrm{d}(A_{\mu \mu }^*A_{\nu \nu } + A_{\mu \nu }^*A_{\nu \mu } 
	- \frac{2}{3} A_{\mu \nu }^*A_{\mu \nu }), \label{dipole-e}& \\
	&f_{\mathrm{field}}=g_\mathrm{m} H_\mu A_{\mu i}^*H_\nu A_{\nu i}. \label{field-e}&
\end{eqnarray}
In the weak-coupling limit, the GL parameters and coupling constant of the dipole energy~\cite{thuneberg} are
\begin{eqnarray*}
	&\alpha = \alpha_0(1-T/T_c),\quad \alpha_0= \frac{N(0)}{3},&\\
	&\beta_2^W = \beta_3^W = \beta_4^W = -\beta_5^W = -2\beta_1^W \\
	&= 2\beta_0^W = \frac{7\zeta (3)N(0)}{120(\pi k_B T_c)^2},&\\
	&K_1 = K_2 = K_3 = \frac{7\zeta (3)N(0)(\hbar v_F)^2}{240(\pi k_B T_c)^2}, &\\
	&g_\mathrm{d}=\frac{\mu _0}{40}\left[ \gamma \hbar N(0) \ln \frac {1.1339 \times 0.45 T_F}{T_c}\right],&\\
	&g_\mathrm{m}=\frac{7\zeta(3)N(0)(\gamma \hbar)^2}{48[(1+F^a_0)\pi k_B T_c]}.&
\end{eqnarray*}
The details of the physical constants in the form described above are as follows: 
the transition temperature $T_c$, the density of states $N(0)$, the Fermi velocity $v_F$, the permeability of vacuum $\mu_0$, 
the gyromagnetic ratio $\gamma$, and the Fermi temperature $T_F$. 
These are given by the experiments~\cite{wheatley, greywall} and depend on the pressure.
In the high-pressure region, the GL parameters of the bulk 4th-order terms $\beta_i$ are corrected by the strong-coupling effect
due to spin fluctuations.
For $\beta_i$, we use the strong-coupling correction calculated by Sauls and Serene~\cite{sauls}, as mentioned below.

\section{Energetics of Vortex Textures}\label{energetics}
	Without the loss of generality, we use the description for the OP $A_{\mu i}=d_{\mu }A_{i}$, 
	where the $d$-vector $d_{\mu }$ and $A_{i}$ are complex values.
	For the bulk of the $^3$He-A phase, $d$-vectors can be unit vectors, 
	but we consider the generic situation on $d$-vectors throughout this work.
	First, let us define textures under the situation that all the $d$-vectors lie in the $xy$-plane. 
	This situation is approximately realized in $^3$He between parallel plates under the strong field 
	$H \gg H_d$.
	If we choose the direction perpendicular to the plane as the spin quantization axis,
	then $A_{\uparrow \downarrow, m}=0$.
	In this case, we can find the following two possibilities on the vortex textures.
	One of them is the HQV: In the vortex core, either the $A_{\uparrow \uparrow, m}$ or $A_{\downarrow \downarrow, m}$
	component of the OP has a unit winding number and singularity.
	The other is the SV: The singularities of both the spin components of OPs are in the same position.
	
	In the zero-field case,
	a single SV has two zero energy modes originating from the spin $\left|\uparrow \uparrow \right>$ 
	and $\left|\downarrow \downarrow \right>$ sectors,
	whereas the HQV has a single zero energy mode from the spin sector that has a phase singularity in the OP.
	The low-energy bound state at the vortex core of the HQV is the same as that of the singular vortex 
	in spinless chiral $p$-wave superfluids.
	Therefore, it is well-known that HQVs obey the non-Abelian statistics~\cite{ivanov}.
	
	However, the energetics of the HQV against the SV still remains a problem.
	Specifically, for the realistic set of the GL parameters in $^3$He, 
	the FL correction maintains the stability of the HQV.
	In contrast, it is demonstrated here that the strong-coupling effect on the bulk 4th-order terms 
	that stabilizes the A phase against the B-phase is not favorable for the stability of the HQV.
	On the basis of the GL theory, we examine here the stability of the HQV as a consequence of the competition between the strong-coupling effect and the FL correction.
	
	The strong-coupling effect in the bulk 4th-order term is derived by Anderson and Brinkman~\cite{anderson} as
	\begin{eqnarray}\label{beta-sc}
		\nn \beta_1 = -(1+0.1\delta)\beta_0^W, & \beta_2 = (2+0.2\delta)\beta_0^W, \\
			\beta_3 = (2-0.05\delta)\beta_0^W, & \beta_4 = (2-0.55\delta)\beta_0^W, \\
		\nn \beta_5 = -(2+0.7\delta)\beta_0^W,
	\end{eqnarray}
	where $\delta$ is the spin-fluctuation parameter depending on the pressure.~\cite{vollhardt}
	Thus, the bulk 4th-order term of the GL free energy (\ref{bulk-4-e}) is 
	\begin{eqnarray}\label{bulk-4-e-strong}
		f_{\mathrm{bulk}} ^{(4)} = B_{\mathrm{d}}(|d_{\uparrow \uparrow }|^4 + |d_{\downarrow \downarrow }|^4) 
		+ B_{\mathrm{c}}(|d_{\uparrow \uparrow }|^2|d_{\downarrow \downarrow }|^2),
	\end{eqnarray}
	where
	\begin{eqnarray}
	\nn	B_{\mathrm{d}} = \beta_0^W \left[ (4-0.35\delta)(|A_{+1}|^4+|A_{-1}|^4) \right.\\
		\left.+ (16 -0.55 \delta)|A_{+1}|^2|A_{-1}|^2 \right],
	\end{eqnarray}
	\begin{eqnarray}
	\nn	\label{strong-advantage}	 B_{\mathrm{c}} = - \beta_0^W \delta \left[3.5(|A_{+1}|^4+|A_{-1}|^4) \right.\\
									\left.+ 9|A_{+1}|^2|A_{-1}|^2 \right] < 0,\label{Bc}
	\end{eqnarray}
	where $A_{m }=(A_x - \sgn(m) i A_y)/\sqrt{2}$, $d_{\sigma \sigma }=(d_x - \sgn(\sigma) i d_y)/\sqrt{2}$, 
	and $d_{\uparrow \downarrow } = d_z$.
	If both the spin $\left|\uparrow \uparrow \right>$ and $\left|\downarrow \downarrow \right>$ components of the OP remain finite,
	the free energy decreases with increasing $\delta$
	because $B_\mathrm{c}<0$ in eq.~(\ref{bulk-4-e-strong}).
	The amplitude of the OP in both the spin sectors is enhanced by each other through the strong-coupling effect. 
	In the case of the core of the HQV, one of the spin components must have the singularity 
	so that another component is not enhanced by the mechanism arising from $B_\mathrm{c}<0$.
	
\begin{figure}[tb]
		\begin{center}
			\includegraphics[width=80mm]{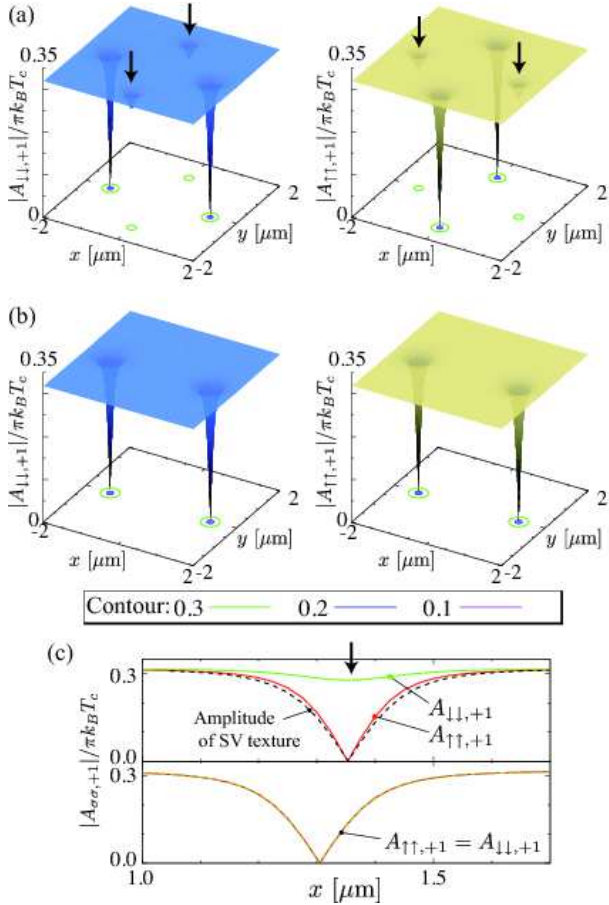}
			\caption{
					(Color online)
					Spatial profiles of the dominant component $|A_{\sigma \sigma, +1}|$ 
					for the four-HQV (a) and the two-SV (b) textures near the center of the system, and 
					(c) their cross section magnifying the vortex core region.
					The upper (lower) panel in (c) is the spatial profile of the HQV (SV).
					The nonsingular component of the HQV is depressed at the vortex core signified by the arrow in (a) and (c).
					The broken line in the upper panel of (c) is the amplitude of $|A_{\sigma \sigma, +1}|$ of the SV 
					when the position of the vortex core coincides with that of the HQV.
					In all the figures, the system size is set to be $R=5$ $\mu$m, the rotating speed is $\Omega= 3\times10^3$ rad/s, 
					and temperature $T=0.95T_c$, 
					where the SV is energetically stable.
					The unit of the $x$- and $y$-axis is micrometers and the amplitudes of the OPs are normalized with $\pi k_BT_c$.
					}
			\label{fig-hqvtex}
		\end{center}
	\end{figure}

	In order to quantify the strong-coupling effect, we carry out the numerical minimization of the GL free energy in eq.~(\ref{glfun}) 
	composed of the bulk, gradient, and dipole energy in eqs.~(\ref{bulk-2-e})-(\ref{dipole-e}).
	Our numerical condition is as follows.
	We assume that the $l$-vectors align to the $z$-direction and the $d$-vectors are in $xy$-plane.
	The uniformity of the OPs is also assumed along the $z$-direction 
	so that the spin and orbital indices of OPs $A_{\mu i} $ reduce to $\mu, i= x,y$ and our calculation is carried out in a 2D plane.
	Then the magnetic interaction energy in eq.~(\ref{field-e}) can be ignored.
	This situation can be realized in a parallel-plate geometry~\cite{yamashita,saunders}, where the distance between the parallel plates is
	less than the dipole coherence length ($\sim 10$ $\mu$m) and much greater than the coherence length ($\sim 10$ nm).
	Here, we impose the rigid boundary condition with the radius $R$ on the OPs as $A_{\mu i}(|\bm{r}|=R)=0$.
	We use the GL parameters $\beta_i$ in eq.~(\ref{bulk-4-e}), taking account of the strong-coupling correction given by Sauls and Serene~\cite{sauls}, 
	and their values are qualitatively consistent with eq.~(\ref{beta-sc}).
	In Fig.~\ref{fig-hqvtex}, we show the amplitudes of the dominant components $A_{\uparrow \uparrow, +1}$ and $A_{\downarrow \downarrow, +1}$ 
	for the SV and HQV.
	In order to compare these energies on an equal footing, we set up two singular vortices for the SV texture and 
	four half-quantum vortices for the HQV texture.
	In our calculation, however, the HQV texture is only the saddle-point solution of the GL equation 
	since the HQV can be continuously transformed into the SV texture and the free energy of the HQV is always higher than that of the SV
	as long as the FL correction is neglected.
	
	In refs.~\ref{salomaahqv}-\ref{vakaryuk}, the OPs in the HQV are restricted within the bulk A phase.
	The strong-coupling effect equally affects the free energy of both the HQV and SV texture under this assumption.
	In our work, we investigate the energetics of the vortices without restrictions even at the vortex core 
	and take account of the strong-coupling effect.
	Thus, the strong-coupling effect near the vortex core gives the HQV relative energetic disadvantage compared with the SV, 
	as shown below.

\begin{figure}[tb]
		\begin{center}
			\includegraphics[width=80mm]{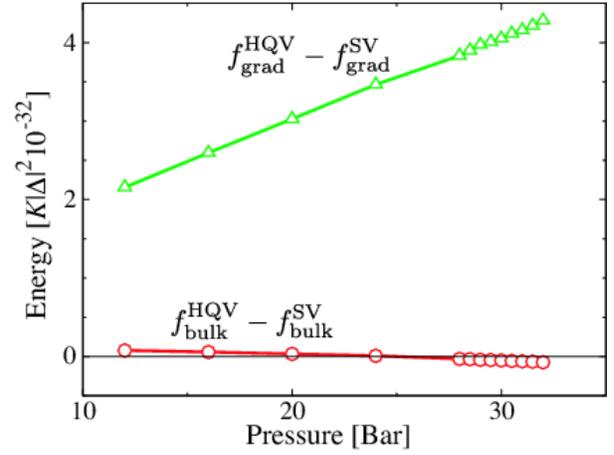}
			\caption{
					(Color online)
					Difference of the gradient and bulk energies of the HQV from that of the SV normalized with $K|\Delta|^2$ as a function of the pressure.
					}
			\label{fig-energetics_gl}
		\end{center}
	\end{figure}

	We analyze the energetics between these two textures in detail.
	In Fig.~\ref{fig-energetics_gl}, we show the difference of the bulk and gradient energies of the HQV from those of the SV, 
	normalized with $K|\Delta|^2$ as a function of the pressure.
	The features of the HQV texture result from two factors: 
	(i) As seen in Fig.~\ref{fig-hqvtex}(c), the amplitude of the component $A_{\downarrow \downarrow, +1}$ is depressed 
	at the singularity of $A_{\uparrow\uparrow, +1}$. 
	(ii) Near the singularity, the component $A_{\uparrow\uparrow, +1}$ in the HQV is more enhanced than that of the SV, 
	as shown in the upper panel of Fig.~\ref{fig-hqvtex}(c),
	because the bulk 4th-order term with $B_\mathrm{c}<0$ in eq.~(\ref{Bc}) makes 
	the interaction between $A_{\uparrow \uparrow, m}$ and $A_{\downarrow \downarrow, m}$ attractive.
	The bulk energy of the $\left|\downarrow \downarrow \right>$ component is increased by the factor (i) 
	and that of the spin $\left|\uparrow \uparrow \right>$ component is reduced by owing to the factor (ii).
	As seen in Fig.~\ref{fig-energetics_gl}, the bulk energies of the HQV relative to the SV is negative in the high-pressure region, 
	and as the pressure decreases, that is, the strong-coupling correction becomes less important, 
	the bulk energy of the HQV eventually becomes higher than that of SV.
	This result implies that the bulk energy with the strong-coupling correction favors the HQV.
	
	However, the gradient energy of the HQV is considerably larger than that of the SV at vortex cores.
	As seen in Fig.~\ref{fig-energetics_gl}, the gradient energy of the HQV relative to that of the SV 
	becomes larger than that of the bulk energy, 
	so that the HQV is not stable with this consideration without the FL correction.
	The reason for this disadvantage of the gradient energy for the stability of the HQV is as follows.
	By the factor (i), that is, the depression of the OP without the vortex singularity at the vortex core of the HQV seen in Fig.~\ref{fig-hqvtex}, 
	the loss of the energy appears owing to the spatial modulation.
	Furthermore, the kinetic energy due to the phase winding is also enhanced by the fact 
	that the singular component $A_{\uparrow \uparrow,+1}$ is enhanced by the factor (ii).
	The strong-coupling effect plays a crucial role in the stability of the HQV in the high-pressure regime, 
	while this becomes less important in the low-pressure regime 
	so that these disadvantages due to the gradient energy may be negligible.
	
	Although the FL correction is the key factor for the stability of the HQV~\cite{salomaahqv, chung, vakaryuk},
	it is difficult to take account of these effects into our calculation on the basis of the GL framework using a more generic form of OPs.
	In order to discuss the FL correction, we apply the London approximation to the OPs, 
	where $A_{+1}(\bm{r})=|\Delta_A|\exp[i\Phi(\bm{r})]$, $A_{-1}(\bm{r})$=0, and the $d$-vector is assumed 
	as the unit vector $[d_x,d_y, d_z]=[\cos\alpha(\bm{r}), \sin\alpha(\bm{r}), 0]$.
	We can treat the FL correction as the effective mass of the spin current using this representation. 
	Note that this representation restricts the pairing phase to the A phase 
	so that we cannot describe the vortex core and boundary of the system exactly.
	The FL correction makes the mass of the spin current less than that of the mass current, that is, $\rho_{\mathrm{sp}}/\rho_{\mathrm{s}}<1$ 
	where $\rho_\mathrm{sp}$ and $\rho_\mathrm{s}$ are the effective masses of the spin and mass currents, respectively.
	
	The gradient term with the FL correction is described by the London approximation as~\cite{vollhardt, salomaahqv}
	\begin{eqnarray}\label{grad-london}
		\frac{f_{\mathrm{grad}}}{2K |\Delta_A|^2 \rho_\mathrm{s}/\rho_\mathrm{s}^0}
		=  (\bm{\nabla}\Phi + \bm{r}\times \Omega)^2 + \left(\rho_\mathrm{sp}/\rho_\mathrm{s}\right)(\bm{\nabla} \alpha )^2.
	\end{eqnarray}
	If we consider an axially symmetric vortex, $\Phi = q^{\mathrm{s}}\phi$, $\alpha = q^{\mathrm{sp}}\phi$, 
	where $\phi$ is the azimuthal angle of the system centered in the vortex singularity and $q^\mathrm{s}$ and $q^\mathrm{sp}$ 
	are the mass and spin circulations, respectively.
	The HQV texture has the circulation $(q^{\mathrm{s}},q^{\mathrm{sp}})=(1/2,1/2)$, whereas the SV has the circulation $(q^{\mathrm{s}},q^{\mathrm{sp}})=(1,0)$.
	Thus, the spatial variation term of the GL free energy for each texture is
	\begin{eqnarray}\label{grad-london-integ}
	\nn	&&\int d \bm{r} \frac{f_{\mathrm{grad}}}{2\pi K|\Delta_A|^2\rho_\mathrm{s}/\rho_\mathrm{s}^0}\\
		&&=\left\{
		\begin{array}{l} 
			 \frac{1}{2}\left(1 +\frac{\rho_\mathrm{sp}}{\rho_\mathrm{s}}\right)  \ln\left(\frac{R}{\xi_{\sigma\sigma }}\right)
			 \ \mathrm{(HQV)}\\ \\
			  \ln\left(\frac{R}{\xi_{\sigma\sigma }}\right)  \ \mathrm{(SV)},
		\end{array}\right.
	\end{eqnarray}
	where the energy of the HQV texture is twice larger than that of the single axially symmetric HQV texture, 
	and we set the external rotation to be $\Omega=0$.
	For $\rho_\mathrm{sp} = \rho_\mathrm{s}$, the energy of the HQV arising from the spatial variation of the OPs is equivalent to that of the SV.
	If $\rho_\mathrm{sp}$ is smaller than $\rho_\mathrm{s}$ owing to the FL correction, the energetic advantage of the HQV texture
	increases logarithmically as the system size $R$ increases.
	
	We numerically estimate this effect in the London limit that neglects the vortex core 
	under the same conditions as those for our full GL calculation, 
	such as the geometry and vortex configuration.
	From the GL calculation, we find that the amplitude of OPs recovers to the bulk within about 0.5 $\mu$m from the vortex core, as shown in Fig.~\ref{fig-hqvtex}.
	Then we neglect this region from the contribution and calculate the spatial variation energy using eq.~(\ref{grad-london}).
	Our numerical calculation is carried out as follows.
	The phase factor is taken as the axially symmetric form for the vortex core $\Phi = q^\mathrm{s}\sum_i \phi_i$, $\alpha = q^\mathrm{sp}\sum_i \phi_i$, 
	where $\phi_i$ is the azimuthal angle for the $j$-th vortex core. 
	The geometry of the system and the layout of the vortices are the same as the GL calculation shown in Fig.~\ref{fig-hqvtex}.
	We set the rotating speed to be $\Omega=\Omega_c$ 
	where the two-SV texture becomes energetically stable against the one-SV texture at the system size $R$. 
	For example, when the system size $R=100$ $\mu$m, the critical speed is found to be $\Omega_c=7.5$ rad/s which is the feasible rotating speed 
	in experiments using a rotating cryostat in ISSP~\cite{yamashita}.
	The integral on the left-hand side of eq.~(\ref{grad-london-integ}) is carried out numerically, 
	and the distance of the vortices is determined by minimizing the gradient energy.
	This distance is consistent with the calculation based on the full GL theory.
	Our model is so simple that it is sufficient for our purpose to estimate only the energy scale of the advantage of the HQV.
	
\begin{figure}[tb]
		\begin{center}
			\includegraphics[width=80mm]{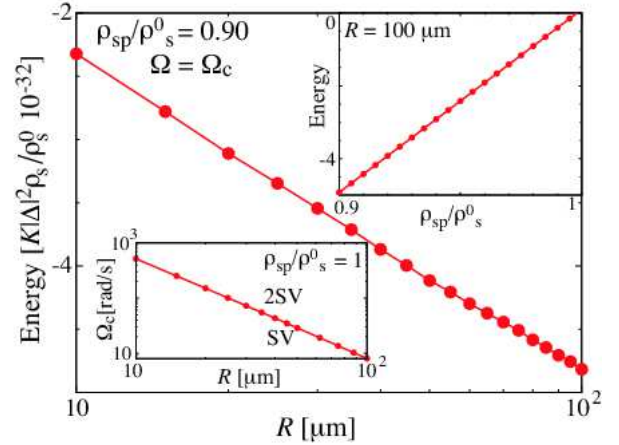}
			\caption{
					(Color online)
					The main panel is the difference of the spatial variation energy $f_\mathrm{grad}$ normalized 
					with $K |\Delta |^2 \rho_\mathrm{s} / \rho_\mathrm{s}^0 $ as a function of the system size $R$
					where the FL correction is set to be $\rho_\mathrm{sp}/\rho_\mathrm{s}=0.9$.
					The rotating speed is set to be $\Omega=\Omega_c$ where the two-SV texture becomes stable compared 
					with the one-SV texture for $\rho_\mathrm{sp}/\rho_\mathrm{s}=1$.
					In our calculation, $\Omega_c$ is determined as a function of the system size shown in the bottom inset.
					The energetic stability of the HQV is proportional to $1-\rho_\mathrm{sp}/\rho_\mathrm{s}$ as seen in the top inset. 
					}
			\label{fig-energetics_london}
		\end{center}
	\end{figure}

	As seen in Fig.~\ref{fig-energetics_london}, the HQV advantage is on the order of $10^{-32}$, normalized with $K|\Delta|^2\rho_\mathrm{s}/\rho_\mathrm{s}^0$.
	By comparing Figs.~\ref{fig-energetics_gl} and \ref{fig-energetics_london}, we find that 
	the relative instability of the HQV originating from the strong-coupling effect becomes comparable to the energetic advantage of 
	the HQV when the FL correction is taken into account.
	
	The strong-coupling correction in the bulk 4th-order terms decreases as the pressure decreases, 
	and the FL effect becomes important as the temperature decreases.
	Hence, the HQV is energetically stable in the lower-pressure and lower-temperature region.
	However, in order to carry out the quantitative calculation of the energetics,
	we need the GL formulation systematically including the strong-coupling bulk 4th-order terms 
	and the FL correction of the gradient terms on an equal footing with the general form of the OPs.
	If we use the most general representation of the OPs, that is, the spin and orbital parts of OPs are not separated, 
	it is impossible to introduce the FL correction by the phenomenological method used by Cross.~\cite{cross}
	These quantitative calculations of the stability remain as a future problem.
	
	As shown above, the HQV has a single ZES bound at the vortex core and obeys the non-Abelian statistics.
	On the other hand, SVs are energetically comparable to the HQV.
	Thus, we consider their statistics in the following section.

\section{Excitations and Braiding of Singular Vortex}\label{SV}
As shown in the previous section, the stability of the HQV remains a problem.
Thus, in this section, we consider the structure of the excitation and braiding of the SV.
In particular, we notice the Zeeman effect due to an external field, and consider the following two situations: 
The magnetic fields applied perpendicular to the $d$-vectors and tilted from its direction.
In this section, we clarify how the spin degrees of freedom of ZESs affect the  statistics of vortices. Here, we only focus on the ZESs and ignore the contributions of the other low-lying core-bound states, because the coupling with the other core-bound states due to thermal fluctuations may give rise to the decoherence of ZESs. This requires the system that we consider here to be in the quantum limit, where the temperature is lower than the energy difference between core-bound states, that is, $T^2_c/T_F$. For $^3$He, since $T^2_c/T_F \!\sim\! 10^{-6}$K, the temperature regime assumed in this section corresponds to the sub-micro Kelvin. This regime does not correspond to the temperature region that we considered within the GL framework in the previous section. However, we consider this limit as a starting point to discuss the statistics of vortices in realistic $^3$He systems. In addition, in the case of ultracold atomic gases with a $p$-wave Feshbach resonance and polar molecules, the superfluid transition temperature $T_c$ is close to $T_F$, because the pair interaction is flexible under an external field. In this case, the braiding operation and the realization of the statistics of vortices are more feasible.

\subsection{$\bm{H}$ perpendicular to $d$-vectors}\label{Hperp}

The excitation composed of the self-Hermitian operators $\gamma _{2j-1}$ and $\gamma_{2j}$ describes the complex fermion eigenstate defined as
\begin{eqnarray}
	c_{2j} = (\gamma_{2j-1} + i\gamma_{2j})/\sqrt{2}.
\end{eqnarray}
The many-body ground state is described by the occupation number of the complex fermion.
The complex fermion state can be understood as a spatially nonlocal state 
when the self-Hermitian excitations $\gamma_{2j-1}$ and $\gamma_{2j}$ are localized at spatially separated vortices.
If the energy eigenvalue of the excitation is exactly zero, an adiabatic braiding of a vortex around another one changes the occupation number.~\cite{ivanov,stern,stone}
For instance, let us consider that there are four vortices $V_i$ ($i=1,2,3,4$) and they have one Majorana quasiparticle $\gamma_i$ at each vortex.
We assume that the Majorana quasiparticles $\gamma_1$ and $\gamma_2$ ($\gamma_3$ and $\gamma_4$) form the complex fermion $c_{2(4)}$.
If the vortex $V_2$ moves around $V_3$, the operator of this braiding is described as~\cite{ivanov,stern} 
\begin{eqnarray}
	\tau_{23} = (c_4^\dagger + c_4)(c_2^\dagger - c_2),
\end{eqnarray}
where we ignore the phase factor.
This operator changes the occupation number of the complex fermion $c_{2i}$.
This feature is known as the non-Abelian statistics of the vortices.
When the odd number of the ZES appears at each vortex core, 
not less than one complex fermion are composed of the ZESs between different vortices.
Then these vortices obey the non-Abelian statistics.

On the other hand, we consider the case where the number of ZESs localized at one vortex core is even.
For instance, the ZESs are spin-degenerate at the core of the well-isolated SV discussed in the previous section.
This leads to the even-number degeneracy of the self-Hermitian operators $\gamma^\uparrow _i$ and $\gamma^\downarrow _i$ 
at the core of each SV $V_i$, arising from the spin $\left|\uparrow \uparrow \right>$ and $\left|\downarrow \downarrow \right>$ sectors.
We consider that there are four vortices.
One can find two possible ways of forming a complex fermion.
(i) The self-Hermitian operators at the same vortex can make the complex fermion described as 
\begin{eqnarray}\label{cf-a}
	a_i=(\gamma_i^\uparrow +i\gamma_i^\downarrow)/\sqrt{2}.
\end{eqnarray}
In this case, an exchange of the vortex means an exchange of the complex Dirac fermion $a_i$, 
implying that the braiding operator gets the only phase factor -1 and belongs to the Abelian group.
(ii) The self-Hermitian operators at the spatially different vortices make the complex fermion described as 
\begin{eqnarray}
	b_{2i}^\sigma=(\gamma_{2i-1}^\sigma + i \gamma_{2i}^\sigma)/\sqrt{2}.
\end{eqnarray}
In this case, if the vortex $V_2$ moves around $V_3$, the braiding operator is described as
\begin{eqnarray}
\nn	\tau_{23} &=& (\{b_4^\uparrow \}^\dagger + b_4^\uparrow)(\{b_2^\uparrow \}^\dagger - b_2^\uparrow)\\
			&&	\times(\{b_4^\downarrow \}^\dagger + b_4^\downarrow)(\{b_2^\downarrow \}^\dagger - b_2^\downarrow).
\end{eqnarray}
When this transformation operates the quasiparticle vacuum state, 
the four complex fermions $b_2^\uparrow$, $b_2^\downarrow$, $b_4^\uparrow$, and $b_4^\downarrow$ are created.
Although the realization of $b_{2i}^\sigma$ is not protected topologically, 
the braiding operator can change the occupation of the complex fermion and the vortices obey the non-Abelian statistics.

The situation where the ZESs are exactly degenerate can have both ways of forming the complex fermions $a_i$ and $b_{2i}^\sigma$.
However, the degeneracy of the ZES is removed by the finite distance of the vortices through quasiparticle tunneling.
We find that the complex fermion state is constructed from the ZESs belonging to different vortices in one of the spin sectors
when the external field is oriented exactly perpendicular to the direction of the $d$-vectors as shown in this section.
We will show that a self-Hermitian particle belonging to a spin state does not hybridize with its counterpart in different spin sectors.
Therefore, the braiding of the vortices changes the occupation number of a complex fermion in each spin sector and the braiding operator is 
approximately non-Abelian.

Under the assumption that the external field $\bm{H}$ is applied to the direction perpendicular to the $d$-vector,
we can block-diagonalize the BdG equation~(\ref{bdg}) into the two spin sectors $\left|\uparrow \uparrow \right>$
and $\left|\downarrow \downarrow \right>$ as
\begin{eqnarray} \label{bdg-block}
\nn \int d\bm{r}_2
	\left[
	\begin{array}{cc}
		\hat{{\mathcal K}}_{\uparrow \uparrow }(\bm{r}_1,\bm{r}_2)  &0  \\
	    0	                                                       & \hat{{\mathcal K}}_{\downarrow \downarrow }(\bm{r}_1,\bm{r}_2)
	\end{array}
	\right]
	\underline{\mathcal{U} }\bm{u}_\nu(\bm{r_2})
	\\=
	E_\nu\underline{\mathcal{U} }\bm{u}_\nu (\bm{r}_1),
\end{eqnarray}
where,
\begin{eqnarray}\label{spinless_k}
	\hat{\mathcal{K}}_{\sigma \sigma }=
	\left[
	\begin{array}{cc}
		H _0 ^\sigma (\bm{r}_1,\bm{r}_2)                 & \Delta _{\sigma \sigma}(\bm{r}_1,\bm{r}_2) \\
		-\Delta _{\sigma \sigma } ^*(\bm{r}_1,\bm{r}_2) & -H_0 ^{\sigma*} (\bm{r}_1,\bm{r}_2)          \\
	\end{array}
	\right],\\
	\underline{\mathcal{U} }\bm{u}_\nu(\bm{r}) = 
	\left[u_{\nu}^{\uparrow}, v_{\nu}^{\uparrow}, u_{\nu}^{\downarrow}, v_{\nu}^{\downarrow}\right]^T,
\end{eqnarray}
where $\underline{\mathcal{U}}$ is an appropriate $4\times 4$ unitary matrix and the spin quantization axis is set to be the direction parallel to $\bm{H}$.
Notice that eq.~(\ref{spinless_k}) is equivalent to the Hamiltonian density of spinless $p$-wave superfluids and 
the components of the wave function belonging to $\left|\uparrow \uparrow (\downarrow \downarrow)\right>$ are given as $(u^{\uparrow(\downarrow)},v^{\uparrow(\downarrow)})$.
In the case of the SV, the two ZESs appear in two spin sectors, 
namely, $\left|\uparrow \uparrow \right>$ and $\left|\downarrow \downarrow \right>$.
We consider systems with a plural number of SVs.
For an infinite vortex distance $D_\mathrm{v}$, the ZES originating from each sector degenerates precisely.
Any linear combination of these two states can be the eigenstate of the system.
However, for a finite $D_\mathrm{v}$, this degeneracy of ZESs can be removed by the following two factors.
One is the Zeeman effect for energy splitting due to quasiparticle tunneling 
and the other is the splitting of the coherence length of two spin components under a strong external field.

First, we discuss the former factor.
In a spinless $p$-wave superfluid, two ZESs bound in neighboring vortices tunnel and interfere with each other.
Then the energy of the complex fermions constructed in an intervortex increases from zero so that the complex fermion states are not exactly ZESs.~\cite{cheng,mizushimaPre}
This energy shift oscillates and decreases exponentially as $D_\mathrm{v}$ increases, 
originating from the quantum oscillation and localization of the wave function of ZESs.

We apply previous spinless argument in the spinful case.
When we take account of the Zeeman effect, the chemical potentials of the up-spin and down-spin particles effectively shift as $\mu\pm \mu_{n}H$,
where $\mu_n$ is the magnetic moment of the particle.
The energy shift of ZESs in the spin $\left|\uparrow \uparrow \right>$ and $\left|\downarrow \downarrow \right>$ sectors can be calculated separately as
\begin{eqnarray}\label{dif-eigenvalue}
\nn	E_{\sigma }&\simeq&-\frac{2|A_{\sigma \sigma,-1}^0|}{\pi ^{3/2}}\left[ \frac{\cos(k_0 D_\mathrm{v}+\pi/4)}{\sqrt{k_0 D_\mathrm{v}}}\right.\\
\nn				 & &\pm\frac{\mu_nH}{\epsilon _F}\frac{k_F}{k_0}\sqrt{D_\mathrm{v}k_0}\sin(D_\mathrm{v}k_0 + \pi/4)\\
				& &+\left. \mathcal{O}\left(\frac{\mu_nH}{\epsilon _F}\right)^2\right]\exp \left(-\frac{D_\mathrm{v}}{\xi_{\sigma\sigma }}\right),
\end{eqnarray}
where $k_0=2M \mu - \xi_{\sigma\sigma }^{-2}$.
Because of the second term in this expression, $E_{\sigma}$, 
the eigenvalues of the spin $\left|\uparrow \uparrow \right>$ sector and $\left|\downarrow \downarrow \right>$ sector 
deviate from each other, and they cannot hybridize with each other.
Thus, complex fermions should be constructed between vortices in each spin sector.

In order to demonstrate this, we numerically diagonalize the spinful BdG equation~(\ref{bdg}). 
This diagonalization is carried out in a two-dimensional system under the OP given in eqs.~(\ref{non-local-op}) and (\ref{gap-potential}).
The system has a circle geometry and the superfluid is confined by the rigid wall potential, as in the GL calculation.
Here, we set $|A_{\sigma \sigma,-1}^0|\ge 0.1\epsilon_F$ in eq.~(\ref{gap-potential}) in order to secure the discreteness of the eigenvalue within the accuracy of calculation.
Here, we consider two vortices that are located at $(x,y)=(-D_\mathrm{v}/2,0)$ and $(D_\mathrm{v}/2,0)$.
Although this calculation is carried out in the strong-coupling region in the sense of $|A_{\sigma\sigma,-1}^0|\sim\epsilon_F$, the features of ZESs are independent of the details of the Hamiltonian.
Therefore, the result of this calculation can be qualitatively applied to the weak-coupling superfluid such as $^3$He.

\begin{figure}[tb]
	\begin{center}
		\includegraphics[width=80mm]{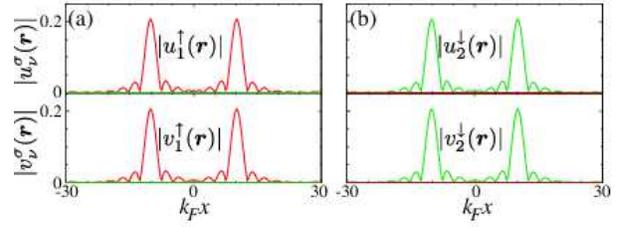}
		\caption{
				(Color online)
				Cross sections of the amplitude of the wave functions $u_1^\uparrow$, $v_1^\uparrow$, $u_2^\downarrow$, and $v_2^\downarrow$ at $y=0$.
				We set the system size $k_FR =30$, the Zeeman splitting $\mu_n H/\epsilon_F = 10^{-4}$, 
				the coherence length of the spin $\left|\sigma \sigma \right>$ component $k_F\xi_{\sigma\sigma } = 2.5$, 
				and the distance of the vortices $k_FD_\mathrm{v} = 20$.
				The panel in (a) shows the wave function of the lowest excitation with $E_1/\epsilon_F=2.08\times10^{-4}$
				and the panel in (b) shows that of the second-lowest excitation with $E_2/\epsilon_F=2.09\times10^{-4}$.
				}
		\label{fig-wfn}
	\end{center}
\end{figure}

\begin{figure}[tb]
	\begin{center}
		\includegraphics[width=80mm]{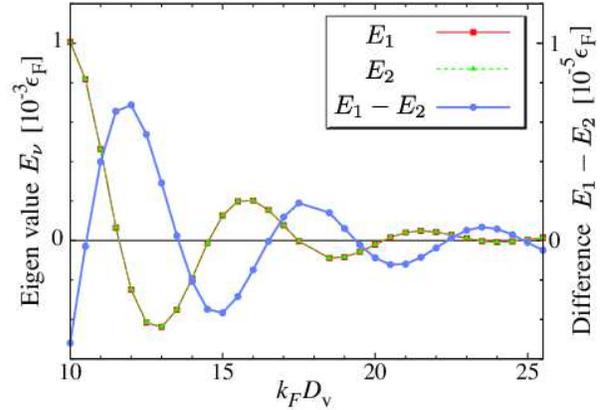}
		\caption{
				(Color online)
				Lowest eigenenergies $E_1$ and $E_2$ originating from 
				the spin $\left|\uparrow \uparrow \right>$ and $\left|\downarrow \downarrow \right>$ sectors 
				and the difference $E_1 - E_2$ as a function of the distance of vortices.
				Here, we set the system size $k_FR =30$, the Zeeman splitting $\mu_n H/\epsilon_F = 10^{-4}$, 
				and the coherence length of the spin $\left|\sigma \sigma \right>$ component $k_F\xi_{\sigma \sigma } = 2.5$. 
				The left and right axes indicate the scale of the eigenenergies $E_\sigma $ 
				and the difference $E_1 -E_2$, respectively.
				}
		\label{fig-ev_oscillation}
	\end{center}
\end{figure}

In Fig.~\ref{fig-wfn}, we show the wave function of the first- and second-lowest excitations
in the presence of Zeeman splitting, $\mu_nH/\epsilon_F = 1.0\times 10^{-4}$, at a distance of the vortices $k_FD_\mathrm{v}=20$.
These excitations are the complex fermion state as a consequence of the tunneling of the ZES bound at vortex cores.
Then near a vortex core, the self-Hermitian relation $u_\nu^\sigma=\{v_\nu^\sigma\}^*$ appears approximately.
As shown in Fig.~\ref{fig-wfn}, these excitations are approximately composed of only one spin component of $u_\nu^\sigma$ and $v_\nu^\sigma$.
The $u_2^\downarrow$ and $v_2^\downarrow$ components of the first-lowest excitation are on the order of $10^{-7}$ of the components $u_1^\uparrow$ and $v_1^\uparrow$.
This means that the eigenstates originating from the $\left|\uparrow \uparrow \right>$ and $\left|\downarrow \downarrow \right>$ sectors are well-separated.
We present in Fig.~\ref{fig-ev_oscillation} the lowest eigenenergies, which oscillate and decay exponentially as a function of $D_\mathrm{v}$.
In Fig.~\ref{fig-ev_oscillation}, the difference between the eigenvalues of the two different spin sectors is found to be about $10^{-2}$ times 
larger than the amplitude of eigenvalues $E_\sigma$, 
and to have the phase difference $\pi/2$ from the eigenvalue oscillation, as shown in eq.~(\ref{dif-eigenvalue}).
In realistic cases, since the amplitude of the OP is much smaller than the Fermi energy 
and the distance of the vortex is larger than this numerical simulation,
these separations of the eigenvalue are quite small but finite.
Therefore, even in a realistic situation, we conclude that complex fermions can be constructed approximately by Majorana ZESs between different vortices in each spin sector.


In the case of spinful chiral $p$-wave superfluids under a strong external field, 
the amplitude and coherence length of the OP component of the spin $\left|\uparrow \uparrow \right>$ pair 
are not equal to those of the $\left|\downarrow \downarrow \right>$ pair.
In the language of $^3$He, this situation is called the A$_2$ phase.
The coherence length $\xi_{\sigma\sigma }$ of the dominant spin component becomes 
smaller than that of the minor component and the ZES is tightly bound at the vortex core. 
The interference through tunneling, shown in eq.~(\ref{dif-eigenvalue}), is weak for the major component and strong for the minor one.
Therefore, the energy difference between the ZESs of each spin sector is enhanced 
by the deviation of $\xi_{\uparrow\uparrow}$ from $\xi_{\downarrow\downarrow}$.
The eigenvalue in the dominant spin sector is closer to zero than that in the minor one.
Thus, the eigenstate in the former has the Majorana character more precisely in the sense that the excitation of the complex fermion state is degenerate with the vacuum state.
The excitation spectrum continuously approaches to that of the spinless case as this imbalance of OPs increases.

\begin{figure}[tb]
	\begin{center}
		\includegraphics[width=80mm]{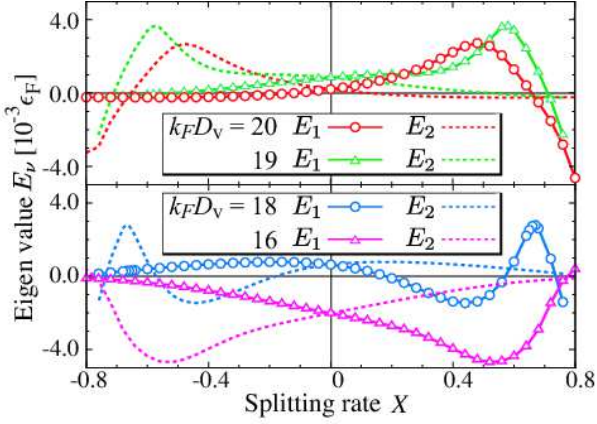}
		\caption{
				(Color online)
				Lowest eigenenergies $E_1 $ and $E_2 $ originating from the spin $\left|\uparrow \uparrow \right>$ (solid line) 
				and $\left|\downarrow \downarrow \right>$ (dashed line) sectors 
				as a function of the splitting rate $X$ defined in eq.~(\ref{splitting}),
				where we set $k_FR =30$, $\mu_n H/\epsilon_F = 10^{-4}$, and $k_FD_\mathrm{v}=16$, $18$, $19$, and $20$.
				The energy differences of the eigenstate originating from two spin sectors (solid and broken line) are 
				on the order of $10^{-6} \epsilon_F$ 
				at the splitting rate $X=0$.
				They are enhanced by increasing $|X|$ to the order of $10^{-3}\epsilon_F$.
				}
		\label{fig-split_uudd}
	\end{center}
\end{figure}

We numerically diagonalize the BdG equation for the OP given as 
\begin{eqnarray}\label{splitting}
\begin{split}
		|A_{\uparrow \uparrow,-1}^0|  = \frac{|\Delta_A|}{\sqrt{2}}(1+X), \\
		\quad |A_{\downarrow \downarrow,-1}^0|= \frac{|\Delta_A|}{\sqrt{2}}(1-X).
\end{split}
\end{eqnarray}
In a realistic system, the splitting $X\in[-1,1]$ is proportional to the external field.~\cite{ambegaokar}
In Fig.~\ref{fig-split_uudd}, we show the behavior of the energy splitting as a function of the imbalance of the OP.
The solid lines are the energy of the eigenstate in the $\left|\downarrow \downarrow \right>$ sector $E_2$ and 
the dashed lines are that of the spin $\left|\uparrow \uparrow \right>$ sector.
In the region $X>0$ ($X<0$), the eigenvalue $E_1$ ($E_2$) 
in the spin $\left|\uparrow \uparrow \right>$ ($\left|\downarrow \downarrow \right>$) sector becomes larger 
and the other eigenvalues $E_2$ ($E_1$) become closer to zero
so that the difference $E_1 - E_2$ also increases.
For instance, at a splitting rate $X=0.48$, this difference is on the order of $10^{-3}\epsilon_F$.
In addition, this energy splitting is more enhanced by the short distance of vortices, as shown in Fig.~\ref{fig-split_uudd}.
The component closer to the zero energy has a Majorana character more precisely because of the degeneracy with the vacuum state of the quasiparticle.

It is not desired that the energy difference between the occupied and vacuum states of the complex fermion is finite.
Because of the energy difference, the adiabatic exchange of vortices cannot change the eigenstate of the complex fermions. 
Thus, we should carry out the braiding for the finite time scale $\epsilon_F/|A_{\sigma\sigma,-1}^0|^2<t<(E_\sigma)^{-1}$,
where the lower boundary depends on the energy discreteness of the core bound states, 
$\Delta E\simeq |A_{\sigma\sigma,-1}^0|^2/\epsilon_F$.~\cite{mizushimaPRA}
In the realistic value of $^3$He, the time scale of $(E_\sigma)^{-1}$ is almost infinite 
and $\epsilon_F/|A_{\sigma\sigma,-1}^0|^2$ is also much larger than the time scale of experiments.
In the case of a $p$-wave resonant Fermi gas~\cite{gurarieAnn,gurariePRB,mizushimaPRL,tsutsumiJPSJ,massignan,mizushimaPRA} or two-dimensional polar fermionic molecules~\cite{bruun, cooper, shi}, 
this range of the time scale may be feasible in experiments.
Furthermore, since there is an energy difference in the many-body ground state between the states before and after the braiding operation, 
the thermal relaxation changes the eigenstate transformed by the braiding.

\subsection{$\bm{H}$ tilted from $\bm{H}\perp$ $d$-vector}\label{Htilt}
In the previous section, we assume that the $d$-vectors are perpendicular to the external field $\bm{H}$.
However, it is difficult to precisely align the $d$-vectors to this direction in experiments.
Thus, we consider the situation where the $d$-vectors are tilted from the direction perpendicular to $\bm{H}$.

Here, we also assume that the $d$-vectors are spatially uniform.
Thus, we choose the directions of the $d$-vectors and the magnetization as $\hat{\bm{d}}=(1,0,0)$ and $\bm{H}=(H_x,0,H_z)$, respectively,
where one finds $A_{\uparrow \uparrow,m} = -A_{\downarrow \downarrow,m }$.

We first consider the eigenstate arising from a well-separated SV and regard the Zeeman effect arising from $H_x$ as perturbation.
In the unperturbed case, the BdG equation is block-diagonalized as shown in eq.~(\ref{bdg-block}),
and the wave functions of the ZESs in these two sectors are described as
\begin{eqnarray}
	\bm{u}_1(\bm{r}) = \left[\ u_1^\uparrow(\bm{r}),\ 0,\ \big\{u_1^\uparrow(\bm{r}) \big\}^*,\ 0\ \right]^T, \\
	\bm{u}_2(\bm{r}) = \left[\ 0,\ u_2^\downarrow(\bm{r}),\ 0,\ \big\{u_2^\downarrow(\bm{r}) \big\}^*\ \right]^T,
\end{eqnarray}
\begin{eqnarray}
	u^\sigma_\nu(\bm{r}) = \exp(i\Phi_\sigma) \mathcal{N} J_0(k^{\sigma}_0r) \exp(-r/\xi_{\sigma\sigma }).
\end{eqnarray}
Here, $\Phi_{\sigma}$ is the phase of OP $A_{\sigma \sigma,-1}$ at the vortex core that arises from the phase of the other vortices, 
$J_0(x)$ is the Bessel function, $\mathcal{N}$ is the normalization constant, and
$k_0^{\sigma }=\sqrt{k_F^2 + \xi_{\sigma\sigma } ^{-2}+\sgn(\sigma) \mu_n H_z/\epsilon_F }$. 
Note that $k_0^\sigma \simeq k_F$ for $k_F\xi_{\sigma\sigma }\gg1$
and $\mu_nH_z\ll\epsilon_F$, and the phase factor $\Phi_\uparrow - \Phi_\downarrow = \pi$ since $A_{\uparrow \uparrow, m} = -A_{\downarrow \downarrow, m}$.
According to the ordinary perturbation theory, the perturbation $H_x$ removes degeneracy as 
\begin{eqnarray}\label{wfn-p}
	\begin{split}
	\bm{u}'_1 = \frac{1}{\sqrt{2}}(\bm{u}_1 + i\bm{u}_2) + \bm{\mathcal{O}}(\mu_n H_x/\epsilon_F), \\
	\bm{u}'_2 = \frac{1}{\sqrt{2}}(\bm{u}_1 - i\bm{u}_2) + \bm{\mathcal{O}}(\mu_n H_x/\epsilon_F). 
	\end{split}
\end{eqnarray}
This wave function means that the operator of the eigenstate under a finite $H_x$ is described by $a_i$ in eq.~(\ref{cf-a}).
Therefore, the two ZESs originating from different spin sectors hybridize in the same vortex core so that the hybridized state behaves as the Dirac fermion.
As shown in the previous section, the vortices that have such a structure of the excitation cannot induce the non-Abelian transformation of the many-body ground state 
by the braiding of the vortices.

\begin{figure}[tb]
	\begin{center}
		\includegraphics[width=80mm]{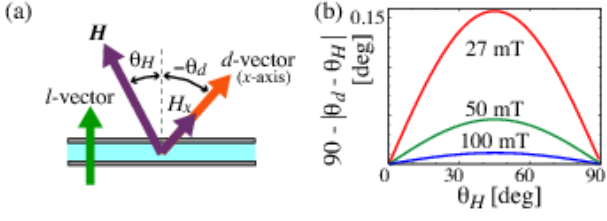}
		\caption{
				(Color online)
				(a) Schematic diagram of the parallel-plate geometry and the definitions of $\theta_H$, $\theta_d$, and $H_x$.
				(b) Angles between the $d$-vector and the external field $\bm{H}$ as a function of $\theta_H$ 
				    calculated in eq.~(\ref{bulkdt}) under the absolute values of the external fields $H=27$, $50$, and $100$ mT.
				}
		\label{fig-dvec_angle}
	\end{center}
\end{figure}

The first-order energy shift due to perturbation is estimated as 
\begin{eqnarray}
\nn	\Delta E^{(1)} &=& \int d\bm{r} \left[\bm{u}'_1(\bm{r})\right]^\dagger 
				\mu_n H_x \underline{\sigma}_x \bm{u}'_1(\bm{r})\\
			&\sim& 4 \mu_n H_x \xi_{\sigma\sigma } k_F, 
\end{eqnarray}
where the $4\times4$ Pauli matrix $\underline{\sigma}_i=\diag[\hat{\sigma}_i,-\hat{\sigma}_i]$.
Here, we use the asymptotic form of the Bessel function $J_0(x)\simeq \sqrt{2/(\pi x)}\cos(x-\pi/4)$ 
and assume that $(\xi_{\sigma\sigma } k_F)^{-1} \ll 1$ and $\xi_{\sigma\sigma } (k_0^\uparrow -k_0^\downarrow) \simeq \xi_{\sigma\sigma } k_F\mu_n H_z/\epsilon _F \ll 1$ using the physical parameters for $^3$He.
In the case of the SV in $^3$He in the parallel-plate geometry defined in Fig.~\ref{fig-dvec_angle}(a),
the angle of the $d$-vectors is determined as 
\begin{eqnarray}
	\theta _d = -\frac{1}{2}\tan ^{-1} \left [\frac{\sin 2\theta _H}{(H_d/H)^2-\cos 2\theta _H}\right], 
	\label{bulkdt}
\end{eqnarray}
where $\theta_d$ and $\theta_H$ are defined in Fig.~\ref{fig-dvec_angle}(a).
Equation~(\ref{bulkdt}) is derived by minimizing the dipole energy in eq.~(\ref{dipole-e}) and the magnetic interaction energy in eq.~(\ref{field-e}) 
within the London approximation, where $A_{+1}(\bm{r})=|\Delta_A|\exp[i\Phi(\bm{r})]$, $A_{-1}(\bm{r})$=0, 
and the $d$-vectors are assumed as the unit vectors. 
The resulting angle $\theta_d-\theta_H$ becomes maximum at $\theta_H=\pi/4$, as shown in Fig.~\ref{fig-dvec_angle}(b).
For instance, using $\theta_H=\pi/4$ and $H=27$ mT, we estimate $H_x=7.40\times 10^{-2}$ mT.
Then, we find $\Delta E^{(1)} = 1.68\times 10^{-8} \ \epsilon _F$, which implies that the shift is much smaller 
than the gap $|\Delta_{\sigma \sigma }|\sim10^{-3}\epsilon_F$ in $^3$He.
However, the symmetry separating two spin sectors is broken by this perturbation, and the complex fermion $a_i$ 
in eq.~(\ref{cf-a}) is constructed in the same vortex core even for an infinitesimal field.

\begin{figure}[tb]
	\begin{center}
		\includegraphics[width=80mm]{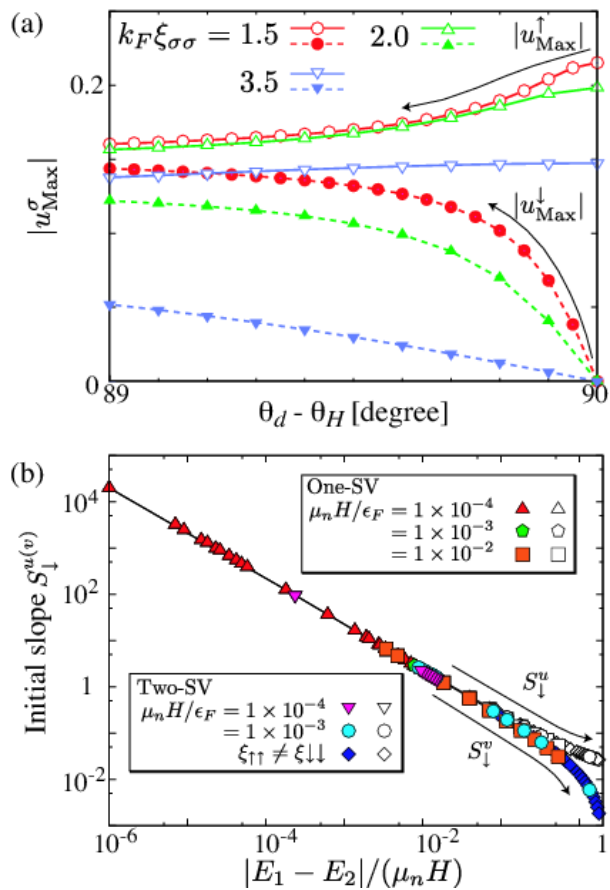}
		\caption{
				(Color online)
				(a) Maximum amplitude $|u_\mathrm{Max}^\sigma|$ of the wave functions $u_1^\sigma(\bm{r})$ 
				as a function of the angle between the $d$-vector and the magnetic field.
				The dashed line is $|u_\mathrm{Max}^\downarrow|$ and the solid line is $|u_\mathrm{Max}^\uparrow|$.
				(b) The absolute values of the initial slopes $S_\downarrow^u$ and $S_\downarrow^v$ of the hybridization of the minor spin components 
				$|u_1^\downarrow|$ (open symbol) and $|v_1^\downarrow|$ (filled symbol), 
				which are defined in eq.~(\ref{islope}),
				as a function of $|E_1 -E_2|/(\mu_nH)$.
				}
		\label{fig-hybridization}
	\end{center}
\end{figure}

Although the perturbation $H_x$ hybridizes the excitations in spin sectors, the interference through the tunneling of the quasiparticle separates 
the spin sectors shown in the previous section.
In order to clarify this conflict, we diagonalize the BdG equation~(\ref{bdg}) where $\theta_d-\theta_H \neq \pi/2$.
In Fig.~\ref{fig-hybridization}(a), we plot the maximum amplitudes of the wave functions $|u_1^\uparrow(\bm{r})|$ and $|u_1^\downarrow(\bm{r})|$ of the lowest-energy eigenstate
as a function of $\theta_d-\theta_H$.
As shown in the previous section, the component $|u^\uparrow_1(\bm{r})|$ is finite and $u^\downarrow_1(\bm{r})=0$ at $\theta_d-\theta_H=\pi/2$, 
implying that this eigenstate originates from the spin $\left|\uparrow \uparrow \right>$ sector.
As seen in Fig.~\ref{fig-hybridization}(a), when $\theta_d-\theta_H$ deviates from $\pi/2$, the minor component $|u_1^\downarrow(\bm{r})|$ grows rapidly, 
implying that the two sectors hybridize with each other. 
In fact, in the case of the coherence length $k_F\xi_{\sigma\sigma }=1.5$, 
the two components of the wave functions become equal at $\theta_d-\theta_H=(89/180)\pi$: $|u_1^\uparrow(\bm{r})|\simeq|u_1^\downarrow(\bm{r})|$.
This is consistent with eq.~(\ref{wfn-p}).
We carry out the calculation under various $\mu_nH$ and $k_F\xi_{\sigma\sigma }$ values.
Then, we find that the magnitude of the Zeeman shift $\mu_n H$ does not change the behavior of the hybridization.
As shown in Fig.~\ref{fig-hybridization}(a), the hybridization for tilting $\theta_d-\theta_H$ weakens 
with increasing coherence length $k_F\xi_{\sigma\sigma }$.
These results imply that the energy splitting at $\theta_d-\theta_H = \pi/2$ normalized with the amplitude of Zeeman splitting $|E_1 - E_2|/(\mu_nH)$ determines the growth rate of the hybridization with respect to $\theta_d-\theta_H$. 
In order to quantify this, we define the initial slope of the hybridization for the $\uparrow(\downarrow)$-dominant mode described as
\begin{eqnarray}\label{islope}
\begin{split}
	S_{\uparrow(\downarrow)}^u=\frac{d}{d(\theta_d-\theta_H)}\left(\frac{|u_\mathrm{Max}^{\downarrow(\uparrow)}|}{|u_\mathrm{Max}^{\uparrow(\downarrow)}|}\right),\\
	S_{\uparrow(\downarrow)}^v=\frac{d}{d(\theta_d-\theta_H)}\left(\frac{|v_\mathrm{Max}^{\downarrow(\uparrow)}|}{|v_\mathrm{Max}^{\uparrow(\downarrow)}|}\right),
\end{split}
\end{eqnarray}
where $|u_\mathrm{Max}^{\sigma}|$ and $|v_\mathrm{Max}^{\sigma}|$ are the maximum values of the lowest-energy wave functions $|u_1^{\sigma}(\bm{r})|$ and $|v_1^{\sigma}(\bm{r})|$, respectively.
In Fig.~\ref{fig-hybridization}(b), we plot the initial slope at $\theta_d-\theta_H = \pi/2$ as a function of $|E_1 - E_2|/(\mu_nH)$.
As shown in Fig.~\ref{fig-hybridization}(b), all the results under different external fields, and the layouts of vortices (one-SV and two-SV case)
are on the same function,
so that we ensure that the initial slope of the hybridization depends only on this ratio and yields a power law behavior 
in the region $|E_1 -E_2|/(\mu_nH)\le10^{-1}$.

In addition, as shown in the previous section, the splitting of the coherence length, $\xi_{\uparrow\uparrow } \neq \xi_{\downarrow\downarrow }$, involves 
the enhancement of the energy difference of the ZESs.
The initial slope decreases owing to this splitting and the result is presented in Fig.~\ref{fig-hybridization}(b) 
in the small splitting region $|E_1 -E_2|/(\mu_nH)\le 1$ corresponding to $X\le10^{-2}$.
With increasing $|X|$ corresponding to $|E_1 -E_2|/(\mu_nH)\ge 1$, the initial slope $S_\sigma^{u(v)}$ deviates from the line 
in Fig.~\ref{fig-hybridization}(b) 
because the initial slope in eq.~(\ref{islope}) is not a proper indicator of the behavior of the hybridization.
This is due to the spatial expansion of the wave function in the minor spin sector.
However in the region $|E_1-E_2|/(\mu_nH)>1$, we find $S_\sigma\le 10^{-1}$. 
Thus, we can control the $d$-vectors to be perpendicular to the magnetic field enough to ignore the hybridization.

One can find that the accuracy of the direction of the magnetic field and the weak coupling of minor component of the pair potentials $A_{\uparrow\uparrow,m}$ or $A_{\downarrow \downarrow,m}$ are required by the hybridization of two spin sectors.
For example, for $^3$He using the rotating cryostat under a high external field~\cite{yamashita}, 
the feasible rotating speed is $\Omega\sim 10$ rad/s, the vortex distance is $D_\mathrm{v}=50$ $\mu$m, and the magnetic field is $H=10$ T.
In this situation, with the tilting angle of the external field $\theta_H\sim 0.10$ deg,
we estimate $\theta_d-\theta_H=10^{-9}$ deg.
It is concluded that even if the coherence length of the minor component of the OP is $\xi_{\sigma\sigma } k_F \ge 10^4$, 
we control the hybridization within $|u_1^\uparrow(r)|/|u_1^\downarrow(r)|\sim 10^{-2}$.

\section{Conclusions}
We have studied the vortices and low-energy excitations of the spinful chiral $p$-wave superfluid 
on the basis of the phenomenological Ginzburg-Landau (GL) theory and the microscopic Bogoliubov-de Gennes theory.
We focus on the $^3$He-A phase between parallel plates under a magnetic field.

In spinful chiral superfluids, possible candidates of the vortex texture are the singular vortex (SV) and the half-quantum vortex (HQV).
In \textsection\ref{energetics}, we have discussed the energetics of these textures.
The free energy of the HQV is decreased by the Fermi liquid (FL) correction~\cite{salomaahqv,chung,vakaryuk}.
However, we find that the strong-coupling correction due to spin fluctuations 
affects the energetics of the vortex texture at the vortex core.
Hence, this correction makes the HQV unstable compared with the SV, 
as a result of our full GL calculation, 
which is not included in the discussion in refs.~\ref{salomaahqv}-\ref{vakaryuk}.
We calculate the contributions of the FL effect using the London approximation and of the strong-coupling effect using the full GL framework separately.
It has been demonstrated that the latter effect becomes comparable with the former under a rotation speed of 5-10 rad/s and near the transition temperature $T_c$, 
which is possible for the experiment for $^3$He using a rotating cryostat in ISSP. 
The quantitative calculation taking account of two effects described above has not been established 
and remains a future problem.

In \textsection4, we have investigated the low-energy excitations and the statistics of the SV in
the quantum limit, where the discreteness of the levels in the vortex core is sufficiently larger than the temperature. For $^3$He, since the energy scale of the discreteness is $T^2_{c}/T_F \!\sim\! 10^{-6}$K, it is difficult to realize the quantum limit in practical experiments. Nevertheless, in this section, we have discussed the statistics of vortices in spinful $p$-wave superfluids and achieved the conclusions summarized below. This may be the starting point for further study of the statistics of vortices in realistic situations that show the decoherence of ZESs and the dissipation of vortex motion. In addition, our results are applicable to vortices in $p$-wave resonant Fermi gases~\cite{gurarieAnn, gurariePRB, mizushimaPRL, tsutsumiJPSJ, massignan, mizushimaPRA} and polar molecules~\cite{bruun, cooper, shi}.
It is well-known that the HQV has the Majorana quasiparticle ZESs, which are topologically protected.
In SVs, the spin $\left|\uparrow \uparrow \right>$ and $\left|\downarrow \downarrow \right>$ components of the order parameter (OP)
have a singularity at the same position 
so that the SV texture has two zero-energy excitations localized at vortex cores.
These behaviors change as the angle between the $d$-vectors and the external field $\bm{H}$ varies.

In \textsection\ref{Hperp}, we have considered the case where the $d$-vectors are exactly perpendicular to $\bm{H}$.
In the case of $^3$He between parallel plates, this situation is realized 
by applying a sufficiently strong external field and sufficiently controlling its direction to be perpendicular to the plates accurately.
In this situation, the SV has the two degenerate ZESs 
that originate from the spin $\left|\uparrow \uparrow \right>$ and $\left|\downarrow \downarrow \right>$ sectors when the vortex distance is infinite.
This leads to the Abelian statistics of vortices.
However, for the finite vortex distance, the ZESs split through the interference of their wave functions.
Then, the degeneracy is removed and the spin $\left|\uparrow \uparrow \right>$ and $\left|\downarrow \downarrow \right>$ sectors cannot hybridize with each other.
Thus, the excitation structures are found to be the same as those in the spinless case, and the SV approximately obeys the non-Abelian statistics.
In addition, when the amplitudes of the OP components split under a high external field, 
that is, the coherence length of the spin $\left|\uparrow \uparrow \right>$ pair is not identical to that of the spin $\left|\downarrow \downarrow \right>$ pair,
the splitting of the eigenenergy is further enhanced by the interference of the zero-energy wave functions.

In \textsection\ref{Htilt}, we have considered the case where the $d$-vectors are tilted from the direction perpendicular to $\bm{H}$.
We have found that when the $d$-vectors are tilted, the eigenstates originating from the spin $\left|\uparrow \uparrow \right>$ and $\left|\downarrow \downarrow \right>$ sectors hybridize intensely and form complex fermions in one of the vortex cores. 
Then our numerical calculation demonstrates that the intensity of the hybridization is determined 
by the energy splitting of the eigenstates of two spin sectors when $d\hbox{-vector}\perp\bm{H}$. 
In order to control this hybridization in experiments, 
one has to carry out the experiment near the spin-polarized state called the A$_2$-phase.

Finally, in the case of $^3$He, we have discussed that the statistics of the vortices in spinful chiral $p$-wave superfluids 
with the vortex distance $D_\mathrm{v}$ depends on external parameters as follows.
In the low-temperature and low-pressure region, the HQV is stable, and the statistics of the vortices is non-Abelian.
In the high-temperature and high-pressure region, the SV is stable as a consequence of the strong-coupling effect. 
When the coherence length of either the spin $\left|\uparrow \uparrow \right>$ or $\left|\downarrow \downarrow \right>$ component of OPs
is much smaller than that on the order of $10^{-1}D_\mathrm{v}$, we cannot control the hybridization of the spin sectors.
Hence, it is found that the statistics of the SVs is Abelian.
However, for coherence lengths larger than $10^{-1}D_\mathrm{v}$, the statistics of the SVs is found to be non-Abelian.
\section*{Acknowledgments}
The authors thank M. Ichioka and Y. Tsutsumi for helpful discussions.
This work was supported by the Japan Society for the Promotion of Science (JSPS) and the "Topological Quantum Phenomena" Grant-in Aid for Scientific Research on Innovative Areas from the Ministry of Education, Culture, Sports, Science and Technology (MEXT) of Japan.


\end{document}